\documentclass[10pt]{article}
\usepackage[utf8]{inputenc}
\usepackage{graphicx, verbatim, url, amssymb, amsfonts, amsthm, latexsym, hyperref, caption, subcaption, rotating}
\usepackage{authblk}
\usepackage{booktabs}
\usepackage[fleqn]{amsmath}
\usepackage{multirow}

\title{Quantifying the higher-order influence of scientific publications}

\author[1]{Massimo Fransceschet}
\affil[1]{\url{massimo.franceschet@uniud.it}. University of Udine, Italy.}
\author[2]{Giovanni Colavizza}
\affil[2]{\url{g.colavizza@uva.nl}. University of Amsterdam, the Netherlands.}

\date{}

\begin{document}

\maketitle

\begin{abstract}
Citation impact is commonly assessed using direct, first-order citation relations. We consider here instead the indirect influence of publications on new publications via citations. We present a novel method to quantify the higher-order citation influence of publications, considering both direct, or first-order, and indirect, or higher-order citations. In particular, we are interested in higher-order citation influence at the level of disciplines. We apply this method to the whole Web of Science data at the level of disciplines. We find that a significant amount of influence -- 42\% -- stems from higher-order citations. Furthermore, we show that higher-order citation influence is helpful to quantify and visualize citation flows among disciplines, and to assess their degree of interdisciplinarity.
\end{abstract}

\section{Introduction}
\label{sec:intro}

New knowledge builds on previous knowledge: this is a central tenet of science. A publication relies on previous publications and cites them to acknowledge this debt \cite{merton_priorities_1957}. Although citations acknowledge direct influences, the extent of the influence of a publication can go beyond these \textit{first-order} relations. The study of the influence of previous publications on new ones rests at the core of scientometrics. The visualization and quantification of such dependence has been termed ``algorithmic historiography'' by Eugene Garfield \cite{garfield_1964,garfield_why_2003}. A variety of tools have been developed for the purpose of facilitating such exploration \cite{chen_citespace_2006,van_eck_software_2010,marx_detecting_2014,thor_introducing_2016,van_eck_citnetexplorer:_2014}. Furthermore, previous literature has investigated methods to trace the historical development of science using citations \cite{lucio-arias_main-path_2008,tu_constructing_2016,subelj_intermediacy_2020} and text \cite{gerow_measuring_2018,jurgens-etal-2018-measuring,soni_follow_2019}. Our related goal here is to quantify citation influence, and thus give credit, beyond direct citations. In particular, we aim at understanding the interplay of first and higher-order influence across academic disciplines.

In this contribution we define \textit{higher-order citations} as citations chains of arbitrary length among pairs of publications, and show how the higher-order citation matrix among disciplines can be computed in an iterative and efficient way. Our proposed method is related to the well-known PageRank algorithm \cite{brin_anatomy_1998,franceschet_pagerank:_2011,waltman_pagerank-related_2014}, but it is specifically focused on quantifying higher-order citation influence. We apply this novel definition to the Web of Science dataset between years 2000 and 2016 included (17,932,523 publications and 190,550,206 citations among them). We show that the contribution of first-order (length 1) citations accounts for 58\% of the whole higher-order citation flow, hence it misses a conspicuous part (42\%) of citation information. Indeed, higher-order citations bring a clear picture of the relationships among disciplines \cite{klavans_toward_2009}. Furthermore, we observe this added value by clustering disciplines into larger communities, finding disciplines that act as brokers among communities, and distinguishing between interdisciplinary and autarchic disciplines. 

\section{Methodology}
\label{sec:methods}

\begin{figure}
	\begin{center}
		\includegraphics[scale=0.35, angle=0]{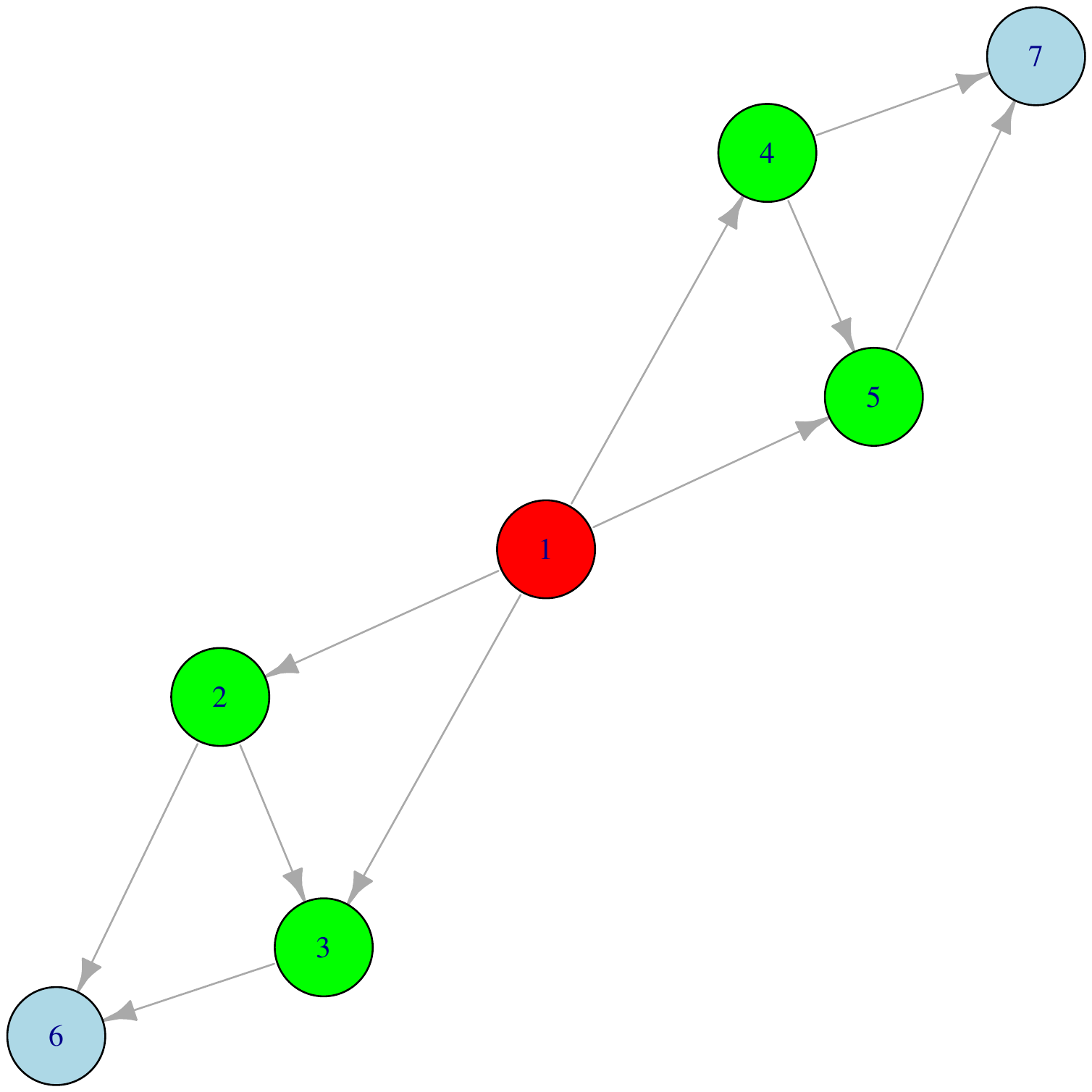}
	\end{center}
	\caption{A publication citation network. The red node is the root (a node with no predecessors) of the DAG, the blue nodes are the leaves (nodes with no successors) and the green nodes are intermediate nodes (vertices with both predecessors and successors).}
	\label{fig:citationNetwork}
\end{figure}

Let $G = (V,E)$ be a citation network with $n$ nodes $V$ and $m$ directed edges $E$. We assume the nodes represent publications. If publication $i$ cites publication $j$, then $(i, j) \in E$. Normally, $G$ is a Directed Acyclic Graph (DAG), because citations only go from more recent publications to older publications.\footnote{There are some exceptions, but these can be removed so as to ensure that $G$ is a DAG.} A simple example is depicted in Figure \ref{fig:citationNetwork}.

Let $A$ be the adjacency matrix, so that $A_{ij} = 1$ whenever $i$ cites $j$, that is $(i, j) \in E$, and $A_{ij} = 0$ otherwise. Let $d_i$ be the outdegree of node $i$, i.e., the number of publications referenced by publication $i$ within the citation network $G$. 

We then recursively define the dependence of publication $i$ on publication $j$ as the mean dependence of publications referenced by $i$ on publication $j$:
$$
P_{ij} =
\begin{cases}
1 & \text{if } i = j,\\
0 & \text{if } i \neq j \text{ and } d_i = 0,\\
\frac{1}{d_i} \sum_k A_{ik} P_{kj} & \text{if } i \neq j \text{ and } d_i > 0.
\end{cases}
$$
We say that $P_{ij}$ is the \emph{dependence} of $i$ on $j$, but on the same note it is the \emph{influence} of $j$ on $i$. Notice that the recursive equation has always a solution since recursion proceeds from each publication to its citing publications, and the graph $G$ is acyclic. 

Let us label each edge of the graph $(i,j)$ with probability $1/d_i$ of going from $i$ to $j$ in a random walk on the graph. Given a path $\pi = k_1, k_2, \ldots k_r$ on the graph, we define the \textit{likelihood} of the path $\pi$ as $$p(\pi) = \prod_{i=1}^{r-1} \frac{1}{k_i}.$$ The dependence $P_{ij}$, when $i \neq j$, is then the sum of likelihoods of all paths from $i$ to $j$ in the graph. In general: 

\begin{quote}
	\em{The dependence $P_{ij}$ is large if there are \textit{numerous likely paths} starting at $i$ and ending in $j$.} 
\end{quote}
For instance, with reference to the graph in Figure \ref{fig:citationNetwork}), we have:

$$
\begin{array}{l}
P_{11} = 1 \vspace{0.15cm} \\
P_{12} = P_{14} = \frac{1}{4} \vspace{0.15cm} \\
P_{13} =  P_{15} = \frac{1}{4} \frac{1}{2} + \frac{1}{4} \frac{1}{1} = \frac{3}{8} \vspace{0.15cm} \\
P_{16} = P_{17} = \frac{1}{4} \frac{1}{2} + \frac{1}{4} \frac{1}{2} \frac{1}{1} + \frac{1}{4} \frac{1}{1} = \frac{1}{2}
\end{array}
$$

We can write this more compactly using matrix notation. Let $D$ be a diagonal matrix such that $D_{ii} = \frac{1}{d_i}$ if $d_i > 0$ and $D_{ii} = 0$ otherwise.
We can then write
\begin{equation} \label{eq:P}
	P = DAP + I
\end{equation}
where $I$ is the $n \times n$ identity matrix. 
We can solve for $P$ and obtain
$$
P = (I - DA)^{-1}
$$

Notice that, if we topologically sort the nodes in $A$ (as done in Figure \ref{fig:citationNetwork}), which is possible since $G$ is a DAG, then both $A$ and $I-DA$ are triangular matrices. In particular, the diagonal elements of $I-DA$ are equal to 1. Hence $\det(I-DA) = 1$, the matrix $I-DA$ is invertible and Equation (\ref{eq:P}) has a solution, as noticed above. The inverse $P = (I - DA)^{-1}$ is also triangular. 

One can also iteratively compute $P$ using the fact that:

\begin{equation}
	P = \sum_{i=0}^{\infty} (DA)^i = \sum_{i=0}^{l} (DA)^i
\end{equation}

where $l \leq n-1$ is the longest path is the graph $G$ and $n$ is the number of nodes of $G$. The last equality holds because $G$ is acyclic and thus $(DA)^i = 0$ for all $i > l$. We expect $l \ll n$. In particular, the length $l$ is bounded by the longest path in the dataset, which corresponds to the number of time instants in the granularity of the dataset. For instance, if the dataset covers 10 years and publication dates are given with a month granularity, then $l$ is lower than $12 \cdot 10 = 120$. 

Matrix $(DA)^i$ computes the dependence contribution of paths of length $i$ in graph $G$. In particular, for $i = 1$, the matrix $DA$ represents \textit{first-order citations}, that is direct citations among publications. On the other hand, matrix $(DA)^i$ for $i > 1$, encodes \textit{higher-order citations}, that is chains of citations of length $i$ among publications. 

Notice that $P_{ij} \neq 0$ if and only if there exists at least on path from $i$ to $j$ in graph $G$. Hence, matrix $P$ has the same non-zero pattern of the adjacency matrix of the transitive closure of $G$. We thus expect $P$ to be denser than $A$. 

\subsection{Discipline dependence}
\label{sec:discipline}

Instead of looking at the individual dependence of publication $i$ on publication $j$, we are interested in disciplinary dependencies. In particular, we are interested in the dependence of a publication (or of a discipline) on a discipline. 

Let us denote by $Q_{iv}$ the extent to which publication $i$ belongs to discipline $v$, hence $Q$ is a matrix $n \times k$, where $n$ is the number of publications and $k$ is the number of disciplines. For the non-overlapping case, $Q_{iv} = 1$ if publication $i$ belongs to discipline $v$. A publication can belong to multiple disciplines, thus $Q_{iv} > 0$ for possibly more than a single discipline $v$. In either case, we have $\sum_{v} Q_{iv} = 1$ and $Q_{iv} \geq 0$.

The dependence $R_{iv}$ of publication $i$ on discipline $v$ can then be defined as the sum of the dependencies of publication $i$ on articles in $v$:
$$
R_{iv} = \sum_j P_{ij} Q_{jv},
$$
or, in matrix notation $$R = P Q.$$
Note that 
$$
\begin{array}{lcl}
R & = & P Q  \\
& = & (DAP  + I) Q \\
& = & DAPQ + Q \\
& = & DAR + Q.
\end{array}
$$

We can hence iteratively compute matrix $R$ without materializing matrix $P$: 

$$
\left\{
\begin{array}{lcl}
R^{(0)} & = & Q  \\
R^{(i+1)} & = & DAR^{(i)} + Q  \\
\end{array}
\right.
$$

Notice that $R^{(i)} = \sum_{j=0}^{i} (DA)^j Q$ is the dependence contribution of citation paths of length up to $i$. Hence $$R = \sum_{i=0}^{\infty} (DA)^i Q = \sum_{i=0}^{l} (DA)^i Q$$ where $l$ is the longest path in the graph, and the iterative computation of $R$ can stop after $l$ steps.  Although $R$ can be as dense as $P$, it has size $n \times k$, which is more manageable than the size of $P$, which is $n \times n$, since we expect $k \ll n$. 

As a particular case, the dependence $r_i$ of publication $i$ on the whole network is $r_i = \sum_j P_{i,j}$, that is, $r = Pe$. We thus have that:
$$r = Pe = (DAP + I)e = DAPe + e = DAr + e.$$
Recall that the Pagerank of $G$, with damping factor $\alpha$ and exogenous vector $\beta$, is the vector $x$ such that $x = \alpha DA x + \beta$ \cite{N10}. Hence, interestingly, the dependence vector $r$ is also the Pagerank of $G$ with damping factor $\alpha = 1$ and exogenous vector $\beta = e$.

One can also define the the dependence $S_{u,j}$ of discipline $u$ on publication $j$ as the sum of the dependence of publications in $u$ on article $j$:
$$
S_{uj} = \sum_i Q_{i u} P_{ij} ,
$$
or, in matrix notation $$S = Q^T P.$$ Notice that since $P = (I - DA)^{-1}$, then $P(I-DA) = I$ and hence $P = PDA+I$. It follows that $S = SDA + Q^T$ and also $S$ can be computed iteratively.

The dependence $F_{uv}$ of discipline $u$ on discipline $v$ is the sum of the dependence of papers in $u$ on papers in $v$, that is:

$$
F_{uv} = \sum_i Q_{iu} R_{iv} = \sum_i \sum_j Q_{iu} P_{ij} Q_{jv},
$$
or, in matrix notation $$F = Q^T R = Q^T P Q = S Q.$$ 
We also define $F^{(i)} = Q^T R^{(i)}$, for $i \geq 0$, as the citation flow matrix for paths of length up to $i$. Notice that, for $i \geq 1$, $F^{(i)} - F^{(i-1)}$ is the citation flow matrix for paths of length equal to $i$. 

\begin{figure}
	\begin{center}
		\includegraphics[scale=0.50, angle=0]{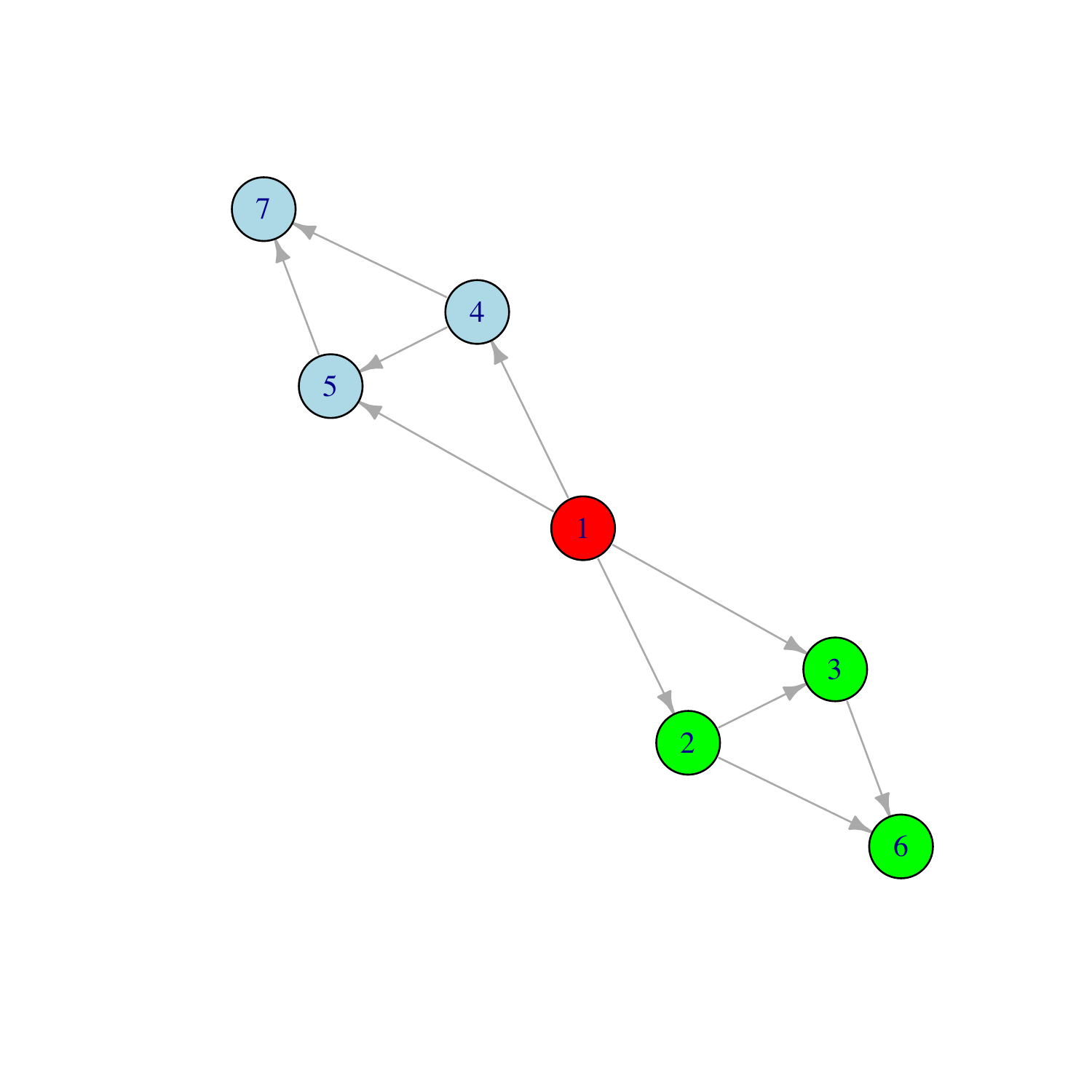}
	\end{center}
	\caption{A publication citation network where nodes are partitioned in 3 non-overlapping disciplines.}
	\label{fig:citationNetwork2}
\end{figure}

Consider again the simple citation network depicted in Figure \ref{fig:citationNetwork2}, where nodes are partitioned in 3 disjoint disciplines. The light blue and green communities are closed worlds (autarchies), since they reference only within their own groups (their off-diagonal flows in matrix $F$ is indeed 0). On the other hand, the red community is more interdisciplinary, since it references the other two groups outside its territory (the off-diagonal flow in matrix $F$ is 2.25).

\section{Case study}

We applied our method on all publications from the CWTS in-house version of the Web of Science, considering the years between 2000 and 2016 included. We consider a total of 17,932,523 publications, and 190,550,206 citations among them -- excluding 444,436 synchronous citations, which we discarded to guarantee that G is a DAG.\footnote{A citation between two publications is discarded if the publication time (year and month) of the citing publication is the same, or older than the publication time of the cited publication.} The longest citation path in the dataset is of length 29 -- equal to the maximum number of iterations needed for convergence. In what follows, we rely on the high-level aggregation of the journal-based classification of Web of Science, which represents 30 broad disciplines (see Table \ref{tab:disciplines}).

\subsection{The contribution of higher-order citations}

We start by assessing the contribution of first-order and higher-order citations to the citation flow among disciplines. Recall that partial flow matrix $F^{(i)}$ is the flow matrix for paths of length up to $i$, with total flow matrix $F = F^{(l)}$, where $l$ is the length of the longest path in the citation graph. Let $M^{(i)} = F^{(i)} - F^{(i-1)}$ be the flow matrix for paths of length precisely $i$. The entry-wise matrix norm $||\cdot||_1$ defined as $||M^{(i)}||_1 = \sum_{u,v} |M^{(i)}_{u,v}|$ is a measure of the total citation flow contained in matrix $M^{(i)}$. We also tested the Frobenius norm $||\cdot||_2$ with similar outcomes. 

\begin{figure}
	\begin{center}
		\includegraphics[scale=0.3, angle=0]{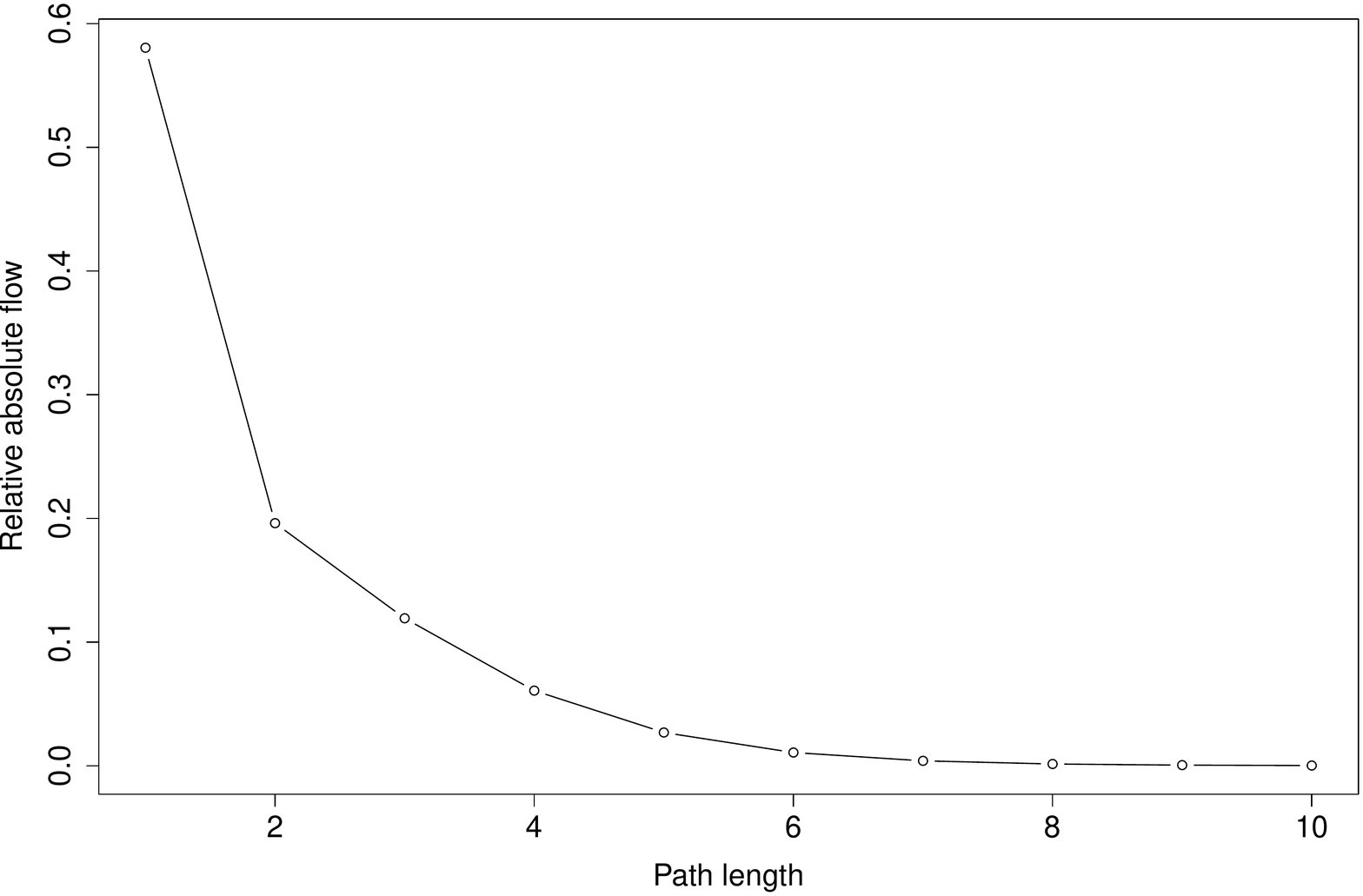}
	\end{center}
	\caption{Relative contribution to the flow of citation paths at given orders (path lengths).}
	\label{fig:relativeFlow}
\end{figure}

\begin{figure}
	\begin{center}
		\includegraphics[scale=0.3, angle=0]{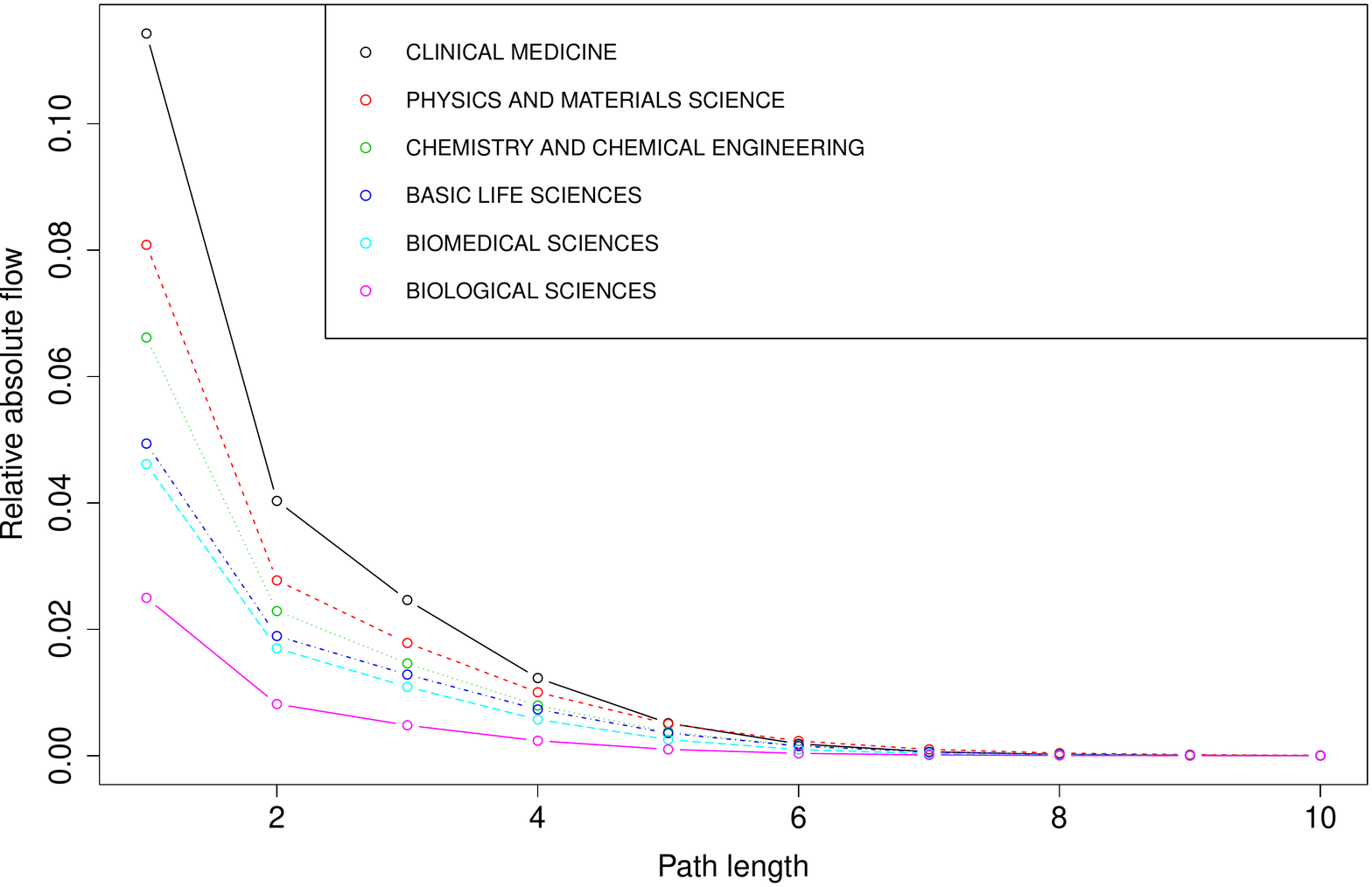}
	\end{center}
	\caption{Relative contribution to the flow of citation paths at given orders (path lengths), for the top 6 disciplines by relative flow contribution.}
	\label{fig:relativeFlowTop}
\end{figure}

We computed the norm of partial flow matrices $M^{(i)}$ relative to the norm of total flow matrix $F = F^{(l)}$, for $1 \leq i \leq l$. Results are shown in Figure \ref{fig:relativeFlow}. First-order (direct) citations contribute for 58\% to the overall flow, hence higher-order citations contribute for 42\%, a significant share. In particular, the share of second-order (length 2) citations is 20\%, that of third-order citations (length 3) is 12\%, and that of fourth-order citations (length 4) is 6\%. Longer citations paths account for about 4\% of the flow. When we consider the top disciplines by flow contribution (Figure~\ref{fig:relativeFlowTop}), we have that six of them account for 38\% (over 42\%) of first-order flow, 13\% (over 20\%) of second-order flow, 8\% (over 12\%) of third-order flow, and 4\% (over 4\%) of fourth-order flow, following a similar pattern to global contributions.\footnote{In order: Clinical medicine, Physics and materials science, Chemistry and chemical engineering, Basic life sciences, Biomedical sciences, Biological sciences.} We conclude that there is an important part of dependence flow that goes beyond direct citations which is worth investigating. 


\subsection{The citation flow network}

\begin{figure}
	\begin{center}
		\includegraphics[scale=0.45, angle=0]{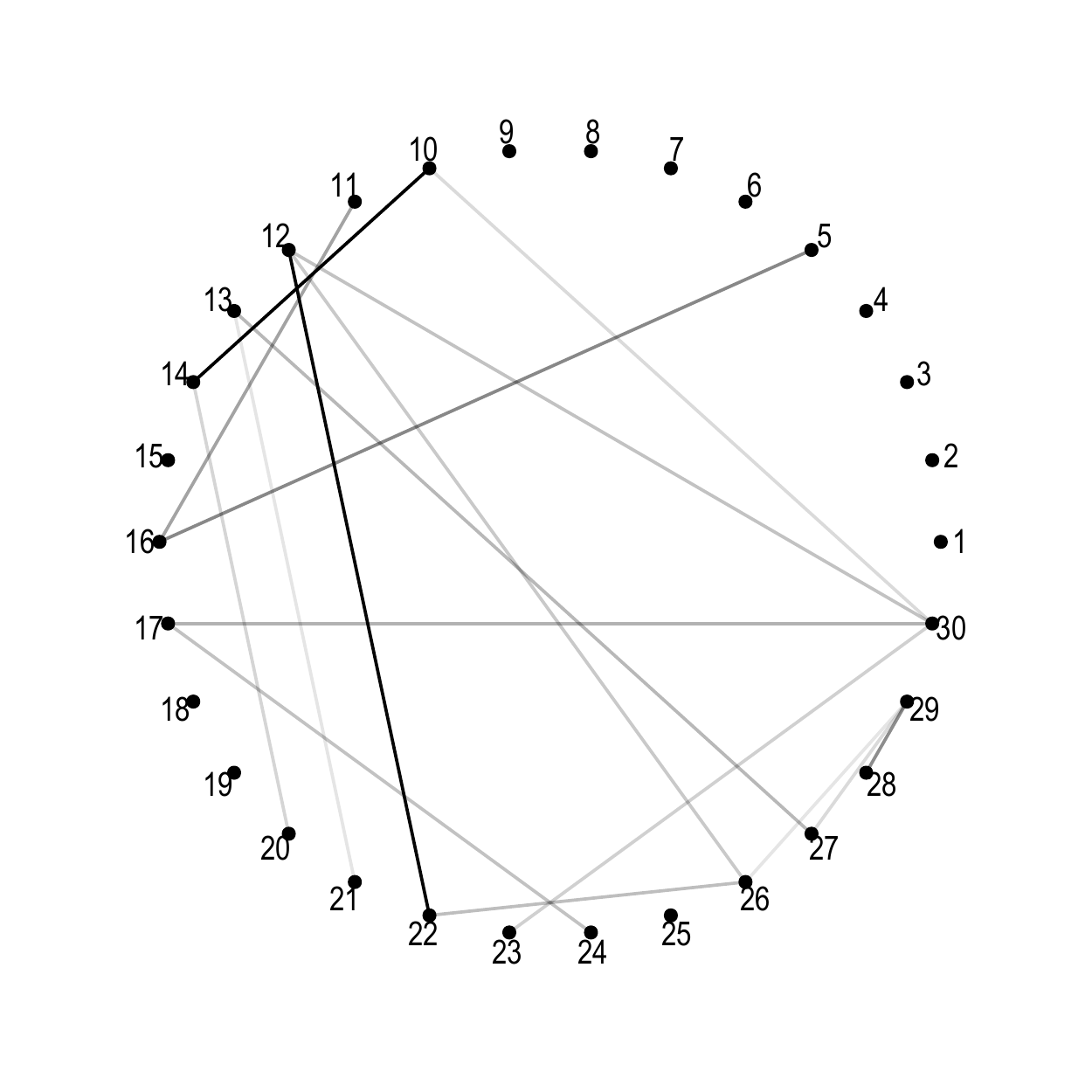}
		\includegraphics[scale=0.45, angle=0]{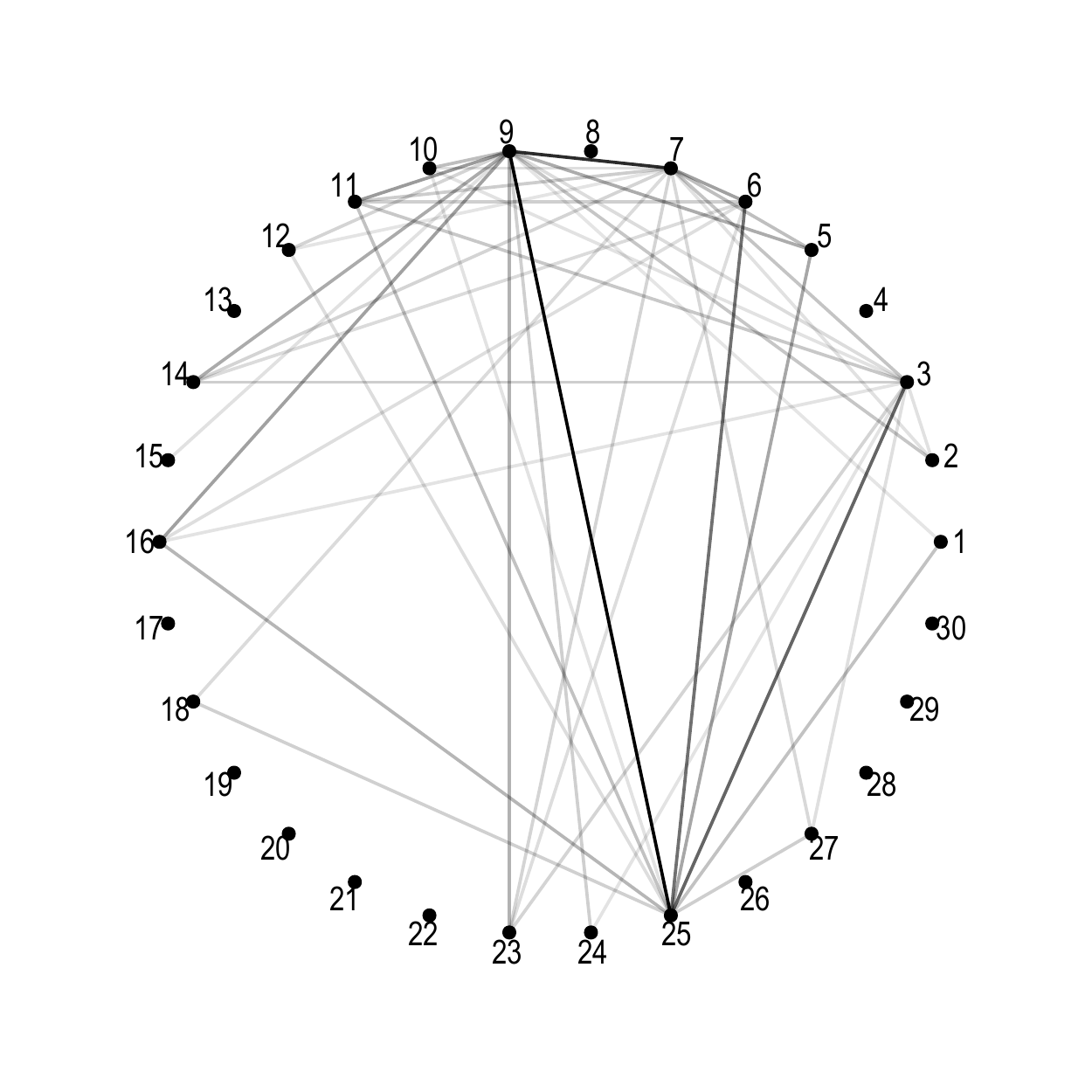}
	\end{center}
	\caption{The higher than expected flows (left) and lower than expected flows (right) among disciplines. E.g., Computer Sciences (10) and Electrical Engineering and Telecommunication (14), as well as Economics and Business (12) and Management and Planning (22) reference each other more than expected, while disciplines Clinical Medicine (9) and Physics and Materials Science (25) reference each other less than expected. See Table \ref{tab:disciplines} for the names of the disciplines.}
	\label{fig:morelessFlow}
\end{figure}

The citation flow matrix is a full matrix and hence the corresponding flow network is a full graph. However, one might investigate the pairs of disciplines that have an higher than expected citation flow, and those that have a lower than expected citation flow. 

Table \ref{tab:disciplines} contains, for each discipline, the internal citation flow (self-flow), the outgoing and incoming citation flows and, moreover, the size of the discipline in number of articles. As expected, citation flows are strongly correlated with size of the discipline (Pearson correlation above 0.9).

To overcome the size-dependence issue, we normalize the flow matrix using the \textit{signed} contribution to Pearson's $\chi$-squared test. The normalized flow $\hat{F}_{i,j}$ between disciplines $i$ and $j$ is computed as: $$\hat{F}_{i,j} = \frac{F_{i,j} - E_{i,j}}{\sqrt{E_{i,j}}}$$ where 
$$
E_{i,j} = \frac{(\sum_k F_{i,k}) \cdot (\sum_k F_{k,j})}{\sum_{u,v} F_{u,v}}
$$ 
is the \textit{expected} flow between $i$ and $j$. The pairs of disciplines that significantly cite each other more than expected (above the 90th percentile) and less than expected (below the 10th percentile) are shown in Figure \ref{fig:morelessFlow}. As for within-discipline citation flows (normalized by expected citations), Astronomy and Astrophysics, Mathematics, and Language and Linguistics lead the ranking, while Instruments and Instrumentation, Basic Medical Sciences and General and Industrial Engineering are at the bottom.  

\begin{figure}
\begin{center}
\includegraphics[scale=0.5, angle=0]{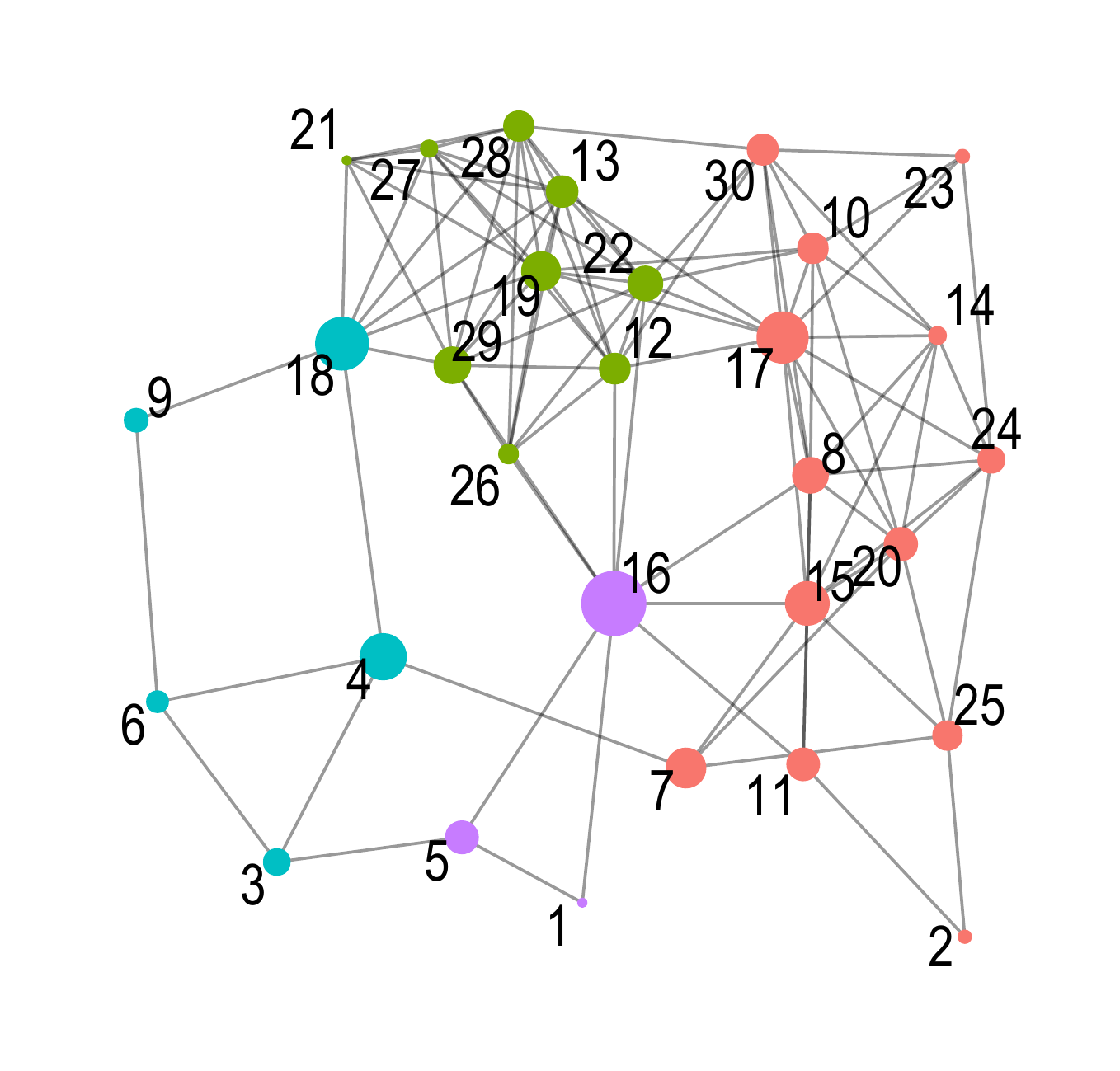}
\end{center}
\caption{The higher-order citation flow network limited to positive (more likely) edges, and divided into communities. Cyan: life and medical sciences; Purple: Earth and environment sciences; Red: mathematical sciences; Green: social and human sciences. We highlight disciplines with large betweenness centrality: Environmental Sciences and Technology (16), Health Sciences (18), and General and Industrial Engineering (17) lead the ranking. See Table \ref{tab:disciplines} for the names of the disciplines. Compare with first-order graph in Figure \ref{fig:community2}.}
\label{fig:community1}
\end{figure}

\begin{figure}
\begin{center}
\includegraphics[scale=0.5, angle=0]{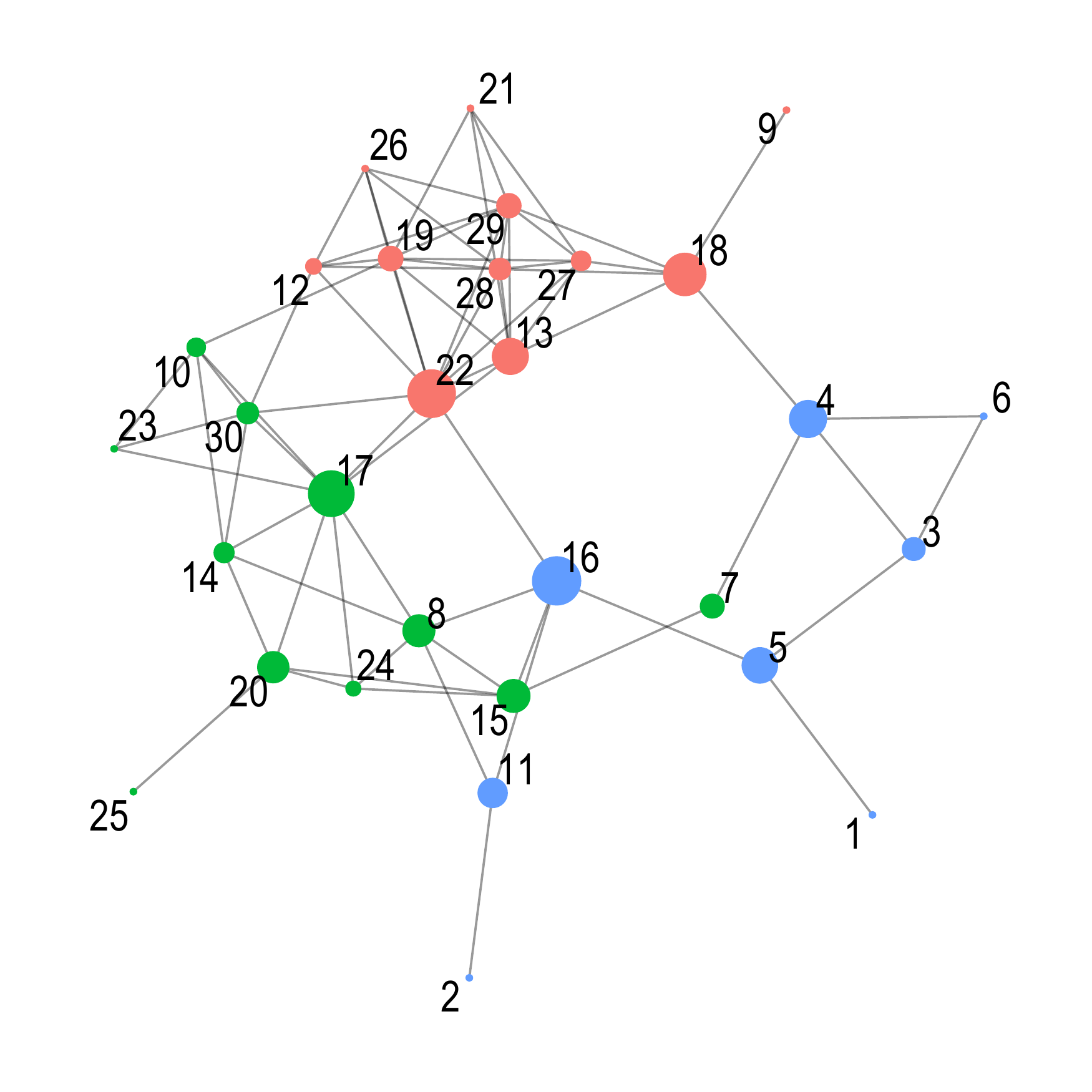}
\end{center}
\caption{The first-order citation flow network limited to positive (more likely) edges, and divided into communities. See Table \ref{tab:disciplines} for the names of the disciplines. Compare with higher-order graph in Figure \ref{fig:community1}.}
\label{fig:community2}
\end{figure}

Furthermore, we consider the same network limited to positively weighted edges, thus with a higher than expected citation flow. We then apply the fast greedy clustering method to this network, as depicted in Figure \ref{fig:community1}. Four macro areas emerge from this analysis, namely the life and medical sciences, science and engineering applied to the Earth and the environment, mathematical sciences and social and human sciences. 
If we do the same limiting ourselves to first-order citations (Figure \ref{fig:community2}), the partition of disciplines into communities is less clear. 

Our analyses suggest that some disciplines are more interdisciplinary (connecting different communities) and other more autarchic (mostly self-referencing), a topic we explore in the following section.


\subsection{Interdisciplinarity and autarchy}

In this section we match higher-order citation flows with measures of interdisciplinarity. We claim that:

\begin{quote}
	\textit{A discipline is interdisciplinary when it is evenly cited from dissimilar disciplines.}
\end{quote}

This thesis immediately recalls the \textit{Rao quadratic entropy} \cite{R82}, which has been previously used to measure interdisciplinarity \cite{porter_is_2009,rafols_diversity_2010,yegros-yegros_does_2015,wang_consistency_2019}. The Rao quadratic entropy is one measure among others which have been studied in the literature~\cite{mugabushaka_bibliometric_2016}. Let us consider a set of objects and a probability distribution $p$ such that $p_i$ is the probability of object $i$. Suppose we also have information about pairwise distance (dissimilarity) $d_{i,j}$ among any two objects $i$ and $j$. Then a measure of heterogeneity among objects is the Rao quadratic entropy:

$$
R(p, d) = \sum_{i,j} p_{i} \, p_{j} \, d_{i,j}
$$

There are two components in this definition of heterogeneity: (1) the evenness of the distribution $p$, (2) the distances $d$ among objects. It holds that, in general: 

\begin{itemize}
	\item $R(p, d)$ is large when $p$ evenly distributes its probability among dissimilar objects;
	\item on the contrary, $R(p, d)$ is small when $p$ concentrates its probability on similar objects.
\end{itemize}

To apply Rao's measure to the higher-order citation flow matrix $F$, we proceed as follows. For each discipline pair $u$ and $v$, let $$p_{u, v} = \frac{F_{u, v}}{\sum_i F_{i, v}}.$$  Notice that $p_{u,v}$ is the relative share of citation flow from discipline $u$ to discipline $v$ compared to the total flow received by $v$. Notice, moreover, that $p_{*, v} = (p_{1, v}, p_{2, v}, \ldots, p_{k, v})$ is a probability distribution.

The similarity $s_{u,v}$ among two disciplines $u$ and $v$ is computed as the cosine of the angle between the $u$ and $v$ columns $F_{*,u}$ and $F_{*,v}$ of the flow matrix $F$:
$$
s_{u,v} = \cos (F_{*,u}, F_{*,v}) = \frac{F_{*,u} F_{*,v}}{\|F_{*,u}\| \|F_{*,v}\|}.
$$  
The cosine runs from 0 (no similarity) to 1 (maximum similarity). Hence, two disciplines are similar if they have a similar pattern of incoming citation flows. The distance $d_{u,v}$ among two disciplines $u$ and $v$ is then $$d_{u,v} = 1 - s_{u,v}.$$ so that two disciplines are distant if they are not similar.

\begin{table}
	\begin{center}
		\begin{tabular}{l|r}
			\hline
			\textbf{Discipline} & \textbf{Rao}\\
			\hline
			Statistical Sciences & 0.678\\
			\hline
			Management And Planning & 0.645\\
			\hline
			General And Industrial Engineering & 0.641\\
			\hline
			Social And Behavioral Sciences, Interdisciplinary & 0.622\\
			\hline
			Civil Engineering And Construction & 0.601 \\ 
			\hline 
			... & ... \\
			\hline 
			Chemistry And Chemical Engineering & 0.360\\
			\hline
			Mathematics & 0.341\\
			\hline
			Astronomy And Astrophysics & 0.316\\
			\hline
			Physics And Materials Science & 0.302\\
			\hline
			Clinical Medicine & 0.294\\
			\hline
	\end{tabular}\end{center}
	\caption{Top 5 (top) and bottom 5 (bottom) disciplines by their interdisciplinarity.}
	\label{tab:id}
\end{table}

Finally, for each discipline $v$, we apply the Rao quadratic entropy to the flow distribution $p_{*, v}$ and distance measure $d$ among disciplines. This gives us a measure of interdisciplinarity for each discipline. The top and bottom 5 interdisciplinary disciplines are given in Table \ref{tab:id}.

Notice how two interrelated disciplines like Statistical Sciences and Mathematics end up on quite different ranks: while Statistics is interdisciplinary, Mathematics is rather autarchic. Indeed, Mathematics receives 78\% of higher-order citation flow from itself, and the rest from a small number of other fields, mainly Physics, Materials Science and Computer Science. On the other hand, the internal flow for Statistics is limited to 43\%. Statistics receives instead a significant citation flow  from many other disciplines, including Mathematics, Computer Sciences, Economics and Business, General and Industrial Engineering, Electrical Engineering and Telecommunication, Clinical Medicine. This suggests that higher-order citations should be considered when assessing the degree of interdisciplinarity or autarchy of a discipline.

\section{Conclusion}
\label{sec:conclusion}

A considerable amount of effort goes into quantifying and assessing citation influence and impact via direct citations. We proposed instead here to quantify citation influence beyond direct citations by also using \textit{higher-order citations}, that is citations chains of arbitrary length among pairs of publications. We have presented a method, informed by PageRank, to quantify the higher-order citation influence of publications. The proposed method accounts for both direct, or first-order, and indirect, or higher-order citations. In particular, we assessed the method on the whole Web of Science corpus between 2000 and 2016 at the level of entire disciplines.

Our results show that the contribution of first-order (length 1) citations accounts for 58\% of the whole higher-order citation flow, while higher-order citations (levels 2 and above) account for 42\%: a significant share. The proposed method is size-dependent, yet easily normalized, and it can be used for a variety of applications. We investigated two here. By using higher-order citation flows, we were able to provide for a high-level map of science clearly distinguishing among four macro-areas: life and medical sciences, Earth and environment sciences, mathematical sciences, social and human sciences. The same picture using only first-order information was found to be less clear-cut. Furthermore, we used the proposed method to rate disciplines according to their degree of interdisciplinarity using the Rao quadratic entropy. We are thus able to distinguish between autarchic disciplines, e.g., mathematics, and interdisciplinary ones, e.g. statistics. We suggest that accounting for higher-order citations is thus relevant and important, and might help on a variety of open scientimetrics questions: performing clustering, measuring interdisciplinarity, assessing the impact of fundamental research, among others.

\section*{Acknowledgements}
\label{sec:ack}

This work stems from prior efforts in collaboration with Ludo Waltman and Vincent A. Traag \cite{colavizza_issi_2019}, whom we thank for their contribution. We are grateful to the Centre for Science and Technology Studies (CWTS), Leiden University, for providing us access to their databases.

\bibliographystyle{plain}
\bibliography{bibliography}

\newpage
\section*{Appendix}

\begin{table}[ht]
\centering
\resizebox{\textwidth}{!}{
\begin{tabular}{r|l|r|r|r|r}
\hline
\textbf{id} & \textbf{discipline} & \textbf{size} & \textbf{self flow} & \textbf{incoming flow} & \textbf{outgoing flow}\\
\hline
1 & agriculture and food science & 875440.50 & 780500.12 & 529167.52 & 743893.92\\
\hline
2 & astronomy and astrophysics & 381254.75 & 686101.56 & 219418.90 & 171588.10\\
\hline
3 & basic life sciences & 2579591.25 & 3456087.00 & 3474212.42 & 2007738.04\\
\hline
4 & basic medical sciences & 268307.25 & 199883.83 & 335008.55 & 483618.99\\
\hline
5 & biological sciences & 1402123.00 & 1259296.75 & 910008.50 & 1164499.91\\
\hline
6 & biomedical sciences & 2507916.50 & 2356196.00 & 2470855.73 & 2487821.15\\
\hline
7 & chemistry and chemical engineering & 3510294.25 & 4352712.50 & 1959569.08 & 2466840.06\\
\hline
8 & civil engineering and construction & 160902.86 & 127872.16 & 132699.23 & 155468.25\\
\hline
9 & clinical medicine & 6024741.50 & 8482322.00 & 3270959.40 & 3051526.10\\
\hline
10 & computer sciences & 647474.88 & 668669.81 & 482215.10 & 506644.46\\
\hline
11 & earth sciences and technology & 934568.50 & 1395625.38 & 549727.39 & 443447.22\\
\hline
12 & economics and business & 429852.88 & 526190.56 & 277452.46 & 185736.94\\
\hline
13 & educational sciences & 238509.97 & 212864.89 & 116494.86 & 163714.45\\
\hline
14 & electrical engineering and telecommunication & 842418.88 & 902059.25 & 629718.60 & 612375.60\\
\hline
15 & energy science and technology & 343416.62 & 196160.98 & 263133.26 & 337039.66\\
\hline
16 & environmental sciences and technology & 983358.88 & 1125205.62 & 886273.78 & 1027153.47\\
\hline
17 & general and industrial engineering & 198930.95 & 101423.06 & 163249.88 & 222303.95\\
\hline
18 & health sciences & 496159.94 & 479285.53 & 429532.91 & 612249.92\\
\hline
19 & information and communication sciences & 104181.30 & 79418.53 & 56385.11 & 76125.22\\
\hline
20 & instruments and instrumentation & 154830.81 & 59613.47 & 153544.22 & 185356.41\\
\hline
21 & language and linguistics & 98703.09 & 80662.09 & 24108.05 & 42272.65\\
\hline
22 & management and planning & 156367.38 & 115467.25 & 143213.02 & 145707.05\\
\hline
23 & mathematics & 831350.88 & 1003179.06 & 281315.60 & 334351.78\\
\hline
24 & mechanical engineering and aerospace & 595979.12 & 489624.09 & 386884.15 & 441418.25\\
\hline
25 & physics and materials science & 4089318.25 & 6163358.50 & 2098397.77 & 2250967.82\\
\hline
26 & political science and public administration & 193848.67 & 170155.39 & 83619.01 & 76208.32\\
\hline
27 & psychology & 581770.75 & 617750.44 & 458871.44 & 412155.69\\
\hline
28 & social and behavioral sciences, interdisciplinary & 132240.47 & 74401.98 & 108292.33 & 128959.83\\
\hline
29 & sociology and anthropology & 218277.44 & 172026.30 & 148015.28 & 173080.76\\
\hline
30 & statistical sciences & 222210.95 & 194457.47 & 252692.38 & 184771.96\\
\hline
\end{tabular}}
\caption{The Web of Science disciplines, with fields id, name of discipline, size, self citation flow, incoming citation flow and outgoing citation flow. Note that the size is the sum of article classifications by discipline. An article can belong to multiple disciplines.}
\label{tab:disciplines}
\end{table}

\end{document}